\documentstyle[12pt,psfig,twoside]{article}
     \newcommand{\ds}{\displaystyle}
     
\begin{document}\hbadness=10000
\thispagestyle{empty}
\pagestyle{myheadings}\markboth{J. Letessier,  J. Rafelski and A. Tounsi}
{Hot Hadronic Matter  at 158 A GeV}
\title{Strangeness in Pb--Pb Collisions at 158 A GeV}
\author{$\ $\\
\bf Jean  Letessier$^1$,  Johann Rafelski$^{2}$ {\rm and} Ahmed
Tounsi$^1$\\ $\ $\\
$^1$Laboratoire de Physique Th\'eorique et Hautes Energies\thanks{\em
Unit\'e  associ\'ee au CNRS UA 280.}\\
Universit\'e Paris 7, 2 place Jussieu, F--75251 Cedex 05.\\ 
$^2$Department of Physics, University of Arizona, Tucson, AZ 85721\\ 
}
\date{June, 1997}   
\maketitle 
\vspace{-9.8cm}{In press in {\it Phys. Lett. B}.
\hfill{\bf PAR/LPTHE/97--25}\\ \vspace*{9.3cm}

\begin{abstract}
{\noindent     
We study relative strange particle abundances
measured  in Pb--Pb 158 A GeV  interactions.  The  thermal
and chemical source parameters of these particles 
are determined under  reaction scenario hypothesis
invoking confined and deconfined hadronic matter.
}   
\end{abstract}
\section*{\normalsize\bf Introduction}
The interest in the study of strange particles 
produced in relativistic 
nuclear collisions derives from the  understanding that 
strangeness flavor  arises most rapidly in  the color 
deconfined  phase.  For reviews of the recent status of the field of 
strangeness as tool in study of dense forms of hadronic matter 
we refer to the proceedings of Strangeness'95 \cite{AIP95} and 
Strangeness'96 \cite{S96} meetings, and  pertinent recent reviews 
\cite{SH95,acta96}.

We recall here specifically that 
studies of Sulphur (S) induced reactions at 200 A GeV show
that strangeness yield is considerably enhanced compared to 
appropriately scaled proton (p) induced reactions. Moreover, 
this effect is accompanied by anomalously high
production of strange antibaryons. It has been suggested that these
effects can arise when strangeness is formed by gluon fusion in high 
density  deconfined phase, which disintegrates in a sudden  
hadronization process without reequilibration 
into final state hadrons. Detailed models were developed showing 
that \cite{acta96,analyze2}.
 
The possibility that strange particle anomalies seen in S-induced
reactions are 
arising from the deconfined quark-gluon-plasma  (QGP) phase
has stimulated the continuation of this
research program in the considerably more difficult, 
high particle  multiplicity environment arising in Pb induced 
reactions, which are presently possible at 158 A GeV.
Here we will attempt a first analysis of  Pb--Pb data
which became available in late Spring 1997.
We advance an  interpretation  of these results
in terms of the formation of a deconfined quark-gluon-plasma 
(QGP),  and  a confined hot hadronic gas (HG). 
 Strategies for making this distinction with hadronic probes 
are considered.

This study addresses  seven
ratios of strange and anti strange baryons 
measured by the WA97--collaboration \cite{WA97} and 
two such ratios measured by the NA49--collaboration \cite{NA49B,NA49O}. 
These ratios appear only on first sight to be much like the
earlier S--W and S--S data of experiments WA85 and WA94 
\cite{WA85}. The NA49--collaboration  has also
presented the rapidity and transverse mass spectra of $\Lambda$, 
$\overline{\Lambda}$ and kaons \cite{NA49B}; the latter result 
are consistent with the 
central transverse mass spectra already reported
by the NA44--collaboration \cite{NA44}. First NA49 results 
about the production of $\phi=s\bar s$ are also 
already available \cite{NA49F}.
In the context of a thermal model interpretation
of the data which we will pursue here, we note
 that the experiment NA49 
reports  a major change in the shape of rapidity
spectra \cite{NA49B}: the $\Lambda, \overline{\Lambda}$ and  kaon 
rapidity  distributions are localized  around 
mid-rapidity and are, in particular in case of $\Lambda$, 
narrower than previously seen in 
S-induced reactions, with a rapidity shape now quite similar
to a directly radiating thermal source. 

The WA97--collaboration stresses that at 95\% confidence level
there is at least a factor 2 enhancement in the 
$\Omega/\Xi$ ratio in Pb--Pb compared to p--Pb collisions:
$$
\frac{(\Omega+\overline{\Omega})/(\Xi+\overline{\Xi})|_{\rm Pb-Pb}}
     {(\Omega+\overline{\Omega})/(\Xi+\overline{\Xi})|_{\rm p-Pb\ }}
                       \simeq 3 \ \ (>2)\,.
$$
The significance of the last result is that hadronic cascades tend to 
attenuate the yield of multistrange hadrons in  strangeness
exchange reactions. This can be thus taken as evidence that
direct emission from a hot fireball of deconfined matter is the
prevailing reaction mechanism. 

The NA49--collaboration stresses that there is no major change in total 
specific strangeness  production, compared to the earlier study of 
S--S and S--Au interactions \cite{NA35}. This is in rough agreement with 
our finding within the QGP reaction model \cite{acta96},
since the gluon fusion 
mechanism is sufficiently fast to nearly saturate the available 
phase space in the S and Pb induced reaction systems. For example,
in Figs. 37 and 38 of Ref. \cite{acta96}, we have shown that the 
specific strangeness yield per baryon increases only by about 20\%
for central Pb induced reactions, compared to S-induced reactions.

\section*{\normalsize\bf Strange (Anti)Baryon  Ratios from QGP}
The experiment WA97 \cite{WA97}  has reported several specific 
strange baryon and  antibaryon  ratios from 
Pb--Pb collisions at 158 A GeV, comprising 30\% of inelastic 
interactions. All ratios are obtained in an overlapping kinematic 
window corresponding effectively to
transverse momentum $ p_\bot>0.7$ GeV, within the central 
rapidity region $ y\in y_{cm}\pm0.5$. They have been corrected for weak
interactions cascading  decays. The experimental values are:
\begin{eqnarray}
R_\Lambda=\frac{\overline{\Lambda}}\Lambda = 0.14\pm 0.03\,,  &&\quad
R_\Xi=\frac{\overline{\Xi}}\Xi         = 0.27\pm 0.05\,, \qquad
R_\Omega=\frac{\overline{\Omega}}\Omega   = 0.42\pm 0.12\,,\\
\ \nonumber \\
R_s^p=\frac\Xi\Lambda=0.14\pm 0.02\,,  &&\quad   
R_{\bar s}^p =\frac{\overline\Xi}{\overline\Lambda} = 0.26\pm 0.05\,,\\
\ \nonumber \\
{R'_s}^p=\frac\Omega\Xi =0.19\pm 0.04\,,  &&\quad
{R'_{\bar s}}^p= \frac{\overline\Omega}{\overline\Xi}  = 0.30\pm 0.09\,.
\end{eqnarray}
Here, the lower index $s$,
resp. $\bar s$ reminds us that the ratio measures the density of
strange, resp. antistrange quarks and upper index $p$ that it is done
within a common interval of transverse momenta (and not common
transverse mass).
We should keep in mind
that  only 6 different particle yields were measured and 
thus only 5 ratios are independent of each other. One easily 
finds the two definition constraints:
\begin{eqnarray}
\frac{R_\Xi}{R_\Lambda}=\frac{R_{\bar s}^p}{R_s^p}\,,\qquad
\frac{R_\Omega}{R_\Xi}=\frac{{R'_{\bar s}}^p}{{R'_s}^p}\,.
\end{eqnarray}
However, in our  fit  to these data, we shall retain all seven ratios as
presented and the error $\xi^2$ will correspond to sum of seven relative 
square errors.  On the other hand we keep in mind that in this experiment
alone  there is only 5 independent data points.

There is agreement between WA97 and NA49 on the value of $R_\Lambda$,
even though the data sample of NA49 is taken for more central trigger,
constrained to as few as 4\% of most central  collisions. 
The cuts in $p_\bot$ and $ y$
are nearly identical in both experiments. From Fig. 3 in 
\cite{NA49B} we obtain the value 
$R_\Lambda=0.17\pm0.03$, which we shall combine with the value given by
WA97 and we thus take in out data fit:
\begin{eqnarray}
R_\Lambda=\frac{\overline{\Lambda}}\Lambda = 0.155\pm 0.04\,.
\end{eqnarray}
The experiment NA49 also reported \cite{NA49O}:
\begin{eqnarray}
\frac{\Xi+\overline{\Xi}}{\Lambda+\overline{\Lambda}}=0.13\pm0.03\,.
\end{eqnarray}
While this ratio can be expressed in terms of the three other ratios
\begin{eqnarray}
\frac{\Xi+\overline{\Xi}}{\Lambda+\overline{\Lambda}}=
R_s\frac{1+R_\Xi}{1+R_\Lambda}\,,
\end{eqnarray}
it is again a separate measurement which thus can be fitted independently.

We next introduce all the model parameters used in the
fit of the particle ratios, not all will be required in 
different discussions of the experimental data. For more details
about  the thermo-chemical parameters
we refer to the extensive discussion in the
earlier study of S--S and S--W data 
\cite{acta96,analyze1,analyze,analyze2}: 
\begin{enumerate}
\item $T_f$ is the formation/emission/freeze-out temperature,
depending on the reaction model. 
$T_f$ enters in the fit of abundance ratios of unlike particles
presented within a fixed $p_\bot$ interval. The temperature $T_f$ 
can in first approximation be related 
to the observed high-$m_\bot$ slope $T_\bot$ by:
\begin{equation}\label{shiftT}
T_\bot \simeq T_f \frac{1+v_\bot}{\sqrt{1-v_\bot^2-v_\parallel^2}}\,.
\end{equation}
In the central rapidity region the longitudinal 
flow $v_\parallel\simeq 0$, in order to assure
 symmetry between projectile and target. Thus as long as  $T_f<T_\bot$, 
we shall use Eq.\,(\ref{shiftT}) setting $v_\parallel=0$ 
to estimate  the transverse 
flow velocity $v_\bot$ of the source.
\item $\lambda_q$, the light quark fugacity.
We initially used in our fits both $u,\,d$-flavor fugacities
$\lambda_u$ and $\lambda_d$, but
we determined that the results were 
represented without allowing for up-down quark asymmetry by  the 
geometric average $\lambda_q=\sqrt{\lambda_u\lambda_d}$; moreover the 
fitted up-down quark fugacity asymmetry was found
 as expected in 
our earlier analytical studies \cite{analyze1}.
\item $\lambda_s$, the strange quark fugacity. A source in which
the carriers of $s$ and $\bar s$ quarks are symmetric this parameter
should have the value $\lambda_s\simeq1$, in general in a re-equilibrated 
hadronic matter the value of $\lambda_s$ can be determined requiring
strangeness conservation. 
\item $\gamma_s$, the strange phase space occupancy. Due to 
rapid evolution of dense hadronic matter it is in general highly 
unlikely that the total abundance of strangeness can follow the 
rapid change in the conditions of the source, and thus in general 
the phase space will not be showing an overall abundance equilibrium 
corresponding to the momentary conditions. 
\item We also show when appropriate  in table \ref{t1} 
the parameter $R_C^s$ describing the relative off-equilibrium abundance of
strange mesons and baryons, using thermal equilibrium abundance as reference. 
This parameter is needed, when we have constraint on the strangeness
abundance and/or when  we address the abundance of mesons 
since there is no a priori assurance that
production/emission of strange mesons and baryons should proceed according to 
relative strength expected from thermal equilibrium. Moreover, it is 
obvious that even if reequilibration of particles in hadronic gas should occur, 
this parameter will not easily find its equilibrium value $R_C^s=1$. However, 
due to reactions connecting strange with non-strange particles we 
obtain $R_C^s=R_C$, where $R_C$ is the same ratio  for non-strange mesons 
and baryons, using thermal abundance as reference. The value of $R_C>1$
implies meson excess abundance per baryon, and thus excess specific
entropy production, also expected due to color  deconfinement \cite{entropy}. 
\end{enumerate} 
{\begin{table}[t]
\caption{\small
Values of fitted statistical parameters within thermal model, see text for 
their meaning. Superscript star `*': a fixed input value for equilibrium 
hadronic gas;  subscript `$|c$':
value is result of the imposed strangeness conservation constraint. 
$\chi^2$ is  the total relative square 
error of the fit for  all data points used. First line: Direct emission
QGP model, no meson ratio fit. Second line: same, but with strangeness
conservation yielding $\lambda_s$ and $R_c$ variable. 
Line three: as in line two, with meson 
ratio fitted. Line four: hadronic gas fit.
\label{t1}}
\vspace{0.1cm}
\begin{center}
\begin{tabular}{|c|c|c|c|c|c|} \hline\hline
$T_{\rm f} [MeV]$& $\lambda_{\rm q}$&$\lambda_{\rm s}$&
$\gamma_{\rm s}$&$R_c^s$& $\chi^2$$\vphantom{\ds\frac{\Xi}{\Lambda}}$ \\
\hline\hline
                   272 $\pm$ 74
                 & 1.50 $\pm$ 0.07
                 &   1.14 $\pm$ 0.04
                 &   0.63 $\pm$ 0.10
                 &   ---
                 &   1.0  \\
\hline
                   272 $\pm$ 74
                 & 1.50 $\pm$ 0.08
                 &   1.14$_{|c}$
                 &   0.63 $\pm$ 0.10
                 &   4.21$\pm$ 1.88
                 &   1.0  \\
                     151 $\pm$ 10
                 &   1.54 $\pm$ 0.08
                 &   1.13$_{|c}$
                 &   0.91$\pm$0.09
                 &   0.85$\pm$0.22
                 &     1.5       \\
\hline
                     155 $\pm$ 7\phantom{0}
                &  1.56 $\pm$ 0.09
                &  1.14$_{|c}$
                &  1$^*$
                &  1$^*$
                &   7.6   \\
\hline
\end{tabular}\\
\end{center} 
 
\end{table}}


The relative number of particles of same type 
emitted at a given instance by a hot source is obtained by
noting that the probability to find all the $j$-components contained within 
the $i$-th  emitted particle is
\begin{equation}\label{abund}
N_i\propto \gamma_s^k\prod_{j\in i}\lambda_je^{-E_j/T}\,,
\end{equation}
and we note that the total energy and fugacity of the particle is:
\begin{equation}
E_i=\sum_{j\in i}E_j,\qquad \lambda_i=\prod_{j\in i}\lambda_j\,.
\end{equation}
The strangeness occupancy $\gamma_s$ enters Eq.\,(\ref{abund})
with power $k$,  which equals the number of strange and antistrange quarks in 
the hadron $i$. 
With $E_i=\sqrt{m_i^2+p^2}=\sqrt{m_i^2+p_\bot^2}\cosh y $ 
 we integrate over the transverse momentum range 
as given by the experiment (here $p_\bot>0.6 $ GeV
taking  central rapidity region $y\simeq 0$
to obtain the relative strengths of particles produced. 
We then allow all hadronic resonances to disintegrate 
in order to obtain the final multiplicity of `stable' particles
we require to form the observed ratios. 
This approach allows to compute the relative
strengths of strange (anti)baryons both in case of 
surface emission and equilibrium disintegration of a particle gas since
the phase space occupancies are in both cases properly accounted for by 
Eq.\,(\ref{abund}). The transverse flow phenomena enter in a similar fashion into
particles of comparable mass and cannot impact particle ratios. On the other hand,
the abundance comparison between different kinds of particles (mesons and baryons)
is unreliable in case of emission from unequilibrated source 
and we will only consider this ratio for the case of a disintegrating 
equilibrated gas. Finally, particles that are easily influenced by the
medium, such as $\phi$, require a greater effort than such a
simple model, and are also not explored in depth here.

We obtain the least square fit for the eight above discussed
 (anti)baryon ratios. 
Our first approach is motivated by the reaction picture consisting of
 direct emission from the 
QGP deconfined fireball. The value of statistical parameters controlling
the abundances are thus free of constraints arising in an equilibrated
hadronic gas (HG) state \cite{analyze1}. 
The fitted thermal parameters are presented in the
first line of table \ref{t1} along with the total $\chi^2$ for
the eight  ratios. The fit is quite good, the 
error shown corresponds to the total accumulated error from 8 measurements;
even if one argues 
that it involves 4 parameters to describe 5 truly independent quantities, 
the statistical significance is considerable, considering that 8 different 
measurements are included. We compare the 
experimental and fitted values in second and third 
column of table~\ref{t2}. 
{\begin{table}[ptb]
\caption{\small
Particle ratios: experimental results and different fits; 
column $|c$ refers to  strangeness constrain imposed in hot QGP, 
(the QGP and $|c$ columns are practically the same);
 `cold' column refers  to late particle
formation from QGP with last given ratio also fitted. HG
fit is for fully equilibrated hadronic gas thermal model.
\label{t2}}
\vspace{0.5cm}
\begin{center}
\begin{tabular}{|c|c|ccc|c|} \hline\hline
Ratios$\vphantom{\ds\frac{\Xi}{\Lambda}}$&Experiment
&QGP&$|c$&cold&HG\\
\hline\hline
$\vphantom{\ds\frac{\Xi}{\Lambda}}$${\Xi}/{\Lambda}$ 
&  0.14 $\pm$ 0.02
 &  0.136   &  0.136   &   0.142  &    0.158        \\
$\vphantom{\ds\frac{\Xi}{\Lambda}}$${\bar\Xi}/{\bar\Lambda}$ 
&  0.26 $\pm$ 0.05
  &  0.232   &  0.232  &   0.260   &   0.292        \\
$\vphantom{\ds\frac{\Xi}{\Lambda}}$${\Omega}/{\Xi}$  
&  0.19 $\pm$ 0.04  
  &  0.185      &  0.184     &  0.123   &  0.141       \\
$\vphantom{\ds\frac{\Xi}{\Lambda}}$${\bar\Omega}/{\bar\Xi}$ 
&  0.30 $\pm$ 0.09 
  &  0.320   &  0.320   &  0.228  &  0.262         \\
$\vphantom{\ds\frac{\Xi}{\Lambda}}$${\bar\Lambda}/{\Lambda}$ 
&  0.155 $\pm$ 0.04 
  &  0.153  &  0.153     &  0.141   &  0.131     \\
$\vphantom{\ds\frac{\Xi}{\Lambda}}$${\bar\Xi}/{\Xi}$  
&  0.27 $\pm$ 0.05       
  &  0.261  &  0.260     &  0.259  &  0.242        \\
$\vphantom{\ds\frac{\Xi}{\Lambda}}$${\bar\Omega}/{\Omega}$  
&  0.42 $\pm$ 0.12 
  &  0.451  &  0.451      &  0.480   &  0.450         \\
$\vphantom{\ds\frac{\Xi}{\Lambda}}$
          $(\Xi+\overline{\Xi})/(\Lambda+\overline{\Lambda})$  
&  0.13 $\pm$ 0.03 
  &  0.151  &  0.151      &  0.166 &  0.184            \\
$\vphantom{\ds\frac{\Xi}{\Lambda}}$
          ${\Lambda}/{K^0_s}|_{m_\bot}$  
&  6.2 $\pm$ 1.5 
  &  ---  &  ---      &  5.6  &  5         \\
\hline
\end{tabular}
\end{center}

\end{table}}

The errors seen in table \ref{t1} 
 on the statistical parameters arise in part from strong
correlations among them. In particular
  the very large error in $T_f$ arise from 
the 80\% anti-correlation with $\gamma_s$.  However, some further 
information about the relation of $T_f$ and $\gamma_s$ 
may be garnered from theoretical considerations. We 
evaluate using our dynamical strangeness evolution 
model in QGP \cite{acta96} how the
value of $\gamma_s$ depends on the temperature of particle 
production $T_f$. The most important parameter in such 
a theoretical  evaluation is the initial temperature at which the deconfined
phase is created. We estimated this temperature 
at $T_{\rm in}=320$ MeV \cite{analyze2}.
Further uncertainty of the calculation arises from the strange quark mass 
taken  here  to be $m_s(1\mbox\,GeV)=200$ MeV (the strength of the 
production rate is now sufficiently constrained by the measurement of 
$\alpha_s(M_Z)$).
We choose  a geometric size which comprises a baryon number $B= 300$
 at $\lambda_q\simeq 1.5$, and have verified that
 our result will be little dependent on small
variations in $B$. We show in Fig.\,\ref{gammTf} how the computed
$\gamma_s$ depends
on formation temperature $T_f$. The cross to the right
shows our  fitted value from line 1 in table \ref{t1}. 
It is consistent with the theoretical expectation 
for early formation of the strange (anti)baryons, and implies
that the central fitted values of $T_f$ and $\gamma_s$ can be
trusted, if such a model is presumed. 
\begin{figure}[ptb]
\vspace*{-1.5cm}
\centerline{\hspace*{-.8cm}
\psfig{width=8cm,figure=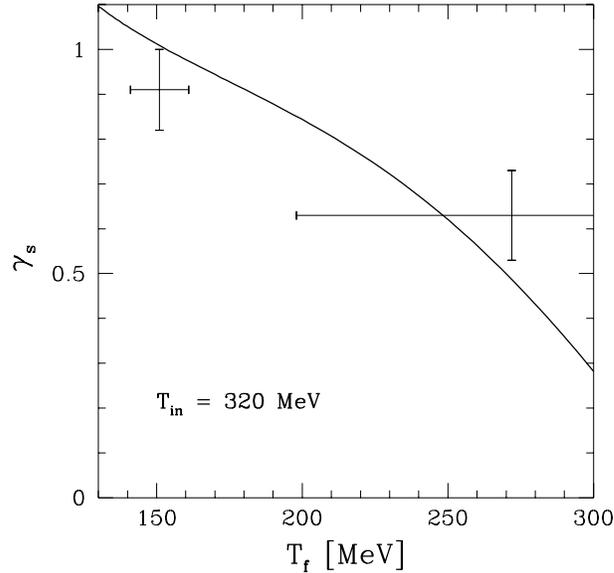}
}
\vspace*{-2cm}
\caption{\small
QGP strangessnes occupancy $\gamma_s$ as function
of temperature $T_f$ at time of particle production, for 
intial temperature $T_{\rm in}=320$ MeV, with 
$\gamma_s(T_{\rm in})=0.1$. See \protect\cite{acta96} for
detaills of the model. 
\label{gammTf}
}

\end{figure}
The relatively high value of temperature $T_f$ we obtained calls for a 
closer inspection of the experimental inverse $m_\bot$  slopes. 
In the common $p_\bot$ range of WA97 and NA49 experiments the transverse 
mass spectrum of $\Lambda$ and $\overline\Lambda$ obtained by NA49 is 
exponential \cite{NA49B}. A thermal model motivated fit the inverse slope 
(temperature) yields $T_\bot^\Lambda=284\pm15$ MeV and 
$T_\bot^{\overline{\Lambda}}=282\pm20$ MeV. This is consistent with
the mid-rapidity proton and antiproton slope of the NA44 experiment: 
$T_p=289\pm7$ MeV and $T_{\overline{p}}=278\pm9$ MeV. For $\Xi+\overline{\Xi}$
a consistent value $T_\Xi=290$ MeV is also quoted by NA49--collaboration~\cite{NA49O}. 
We note that because the baryon masses are large, all 
these slopes are at relatively
high $m_\bot>1.3 $ GeV (for nucleons, in NA44, $m_\bot>1 $ GeV). 
Systematically smaller inverse-transverse slopes    
are seen at smaller $m_\bot$, for Kaons $T_\bot^K\simeq 213$--$224$ MeV 
for $0.7 <m_\bot<1.6$ GeV in NA49 \cite{NA49B} and 
$T_\bot^{K^+}= 234\pm6,\,T_\bot^{K^-}=235\pm7$ MeV in NA44 \cite{NA44};
and 155--185 MeV for $\pi$, \cite{NA49B,NA44},  depending 
on the range of $p_\bot$, but here we have to remember that pions 
arise primarily from resonance decays. An increase of $T$ with $m_\bot$ 
 is most naturally associated with the effects of transverse flow of
the source. In any case we are led to the conclusion that 
$200<T_f<290$ MeV and that as far as the fitted temperatures and slopes are 
concerned it is possible that  the high $m_\bot$ 
strange (anti)baryons we have described could have been emitted directly 
from a primordial (deconfined) phase before it evolves into final 
state hadrons.

Is this hypothesis also consistent with the chemical fugacities 
we have obtained? The chemical condition is 
fixed to about 5\% precision, and there is 40\% anti-correlation
between the two fugacities $\lambda_q$ and $\lambda_s$. The
information that
$\lambda_s\ne1$ is contained in at least two particle
abundances; arbitrary manipulation of the reported yields of one particle 
abundance did not reduce the value $\lambda_s$ to unity. Since $\lambda_s\ne 1$ 
by  4\,s.d.  it is highly unlikely that $\lambda_s=1$ 
is found after more data is studied. While one naively expects 
$\lambda_s^{\rm QGP}=1$, to assure the strangeness balance 
$\langle s-\bar s\rangle=0$, there must be 
a small deviations from this value, even if the emitted particles were to
reach asymptotic distances without any further interactions: 
in presence of baryon density the deconfined state is not
fully symmetric under interchange of particles with antiparticles. 
A possible mechanism to distinguish the strange and anti-strange quarks 
arises akin to the effect considered for the $K^-/K^+$ asymmetry  in baryonic 
matter \cite{Shu92,Go92}: there is asymmetric scattering  strength on 
baryon density $\nu_b$
 which causes presence of a mean effective vector potential
$W$. Similarly, strange quark interaction with baryon density would lead to a
dispersion relation 
\begin{equation}\label{Ws}
E_{s/\bar s}=\sqrt{m_s^2+p^2}\pm W \, ,
\end{equation}
and this requires in the statistical approach that the 
Fermi distribution for strange and antistrange 
quarks acquires a  compensating fugacity $\lambda_{s,\bar s}=e^{\pm W/T}$ 
to assure strangeness balance in the deconfined phase. In linear
response approach $W\propto\nu_b$  consistent with both $W$  and
baryon density $\nu_b=(n_{q}-n_{\bar q)}/3$, 
being fourth component of a Lorenz-vector. 
It is clear for intuitive reasons, as well as given experimental 
observations, that the baryon stopping and thus density increases
considerably comparing the S and Pb induced reactions in the energy domain
here considered. We also recall that in S--W reactions 
 $\lambda_s^{\rm S}\simeq 1.03\pm 0.05$ \cite{analyze}. 
Should in the dense matter fireball the 
baryon density $\nu_b$ grow by factor 2--4 
as the projectile changes from S to Pb, this alone would
consistently explain  the appearance of the value $\lambda_s=1.14\pm0.04$. 
It is worth noting that the value $W\simeq 38$\,MeV suffices here. Note
that the Coulomb potential effect on the charge of the strange quarks is
of opposite magnitude and about 1/5--1/6 of the here required strength.

In order to estimate what would be implied by strangeness conservation 
constraint by the presence of the value $\lambda_s\ne 1$ we present in table 
\ref{t1} the result of a fit assuming that the value of $\lambda_s$ 
is result of the conservation constraint $\langle s-\bar s\rangle= 0$,
and introducing $R_c\ne 1$. The result implies that considerably more
strange mesons would be produced if they were emitted along with baryons
in a sudden and early QGP disintegration. The statistical  error is found
practically the same as in line 1. We note that the third raw in
table \ref{t2} (under heading $|c$ for constrain) is
 presenting the resulting particle ratios for this fit. It is hardly 
changed from the second raw, as the fit converges practically to the same 
point in this approach. 

\section*{\normalsize\bf Late Emission Interpretation}
If particle formation from QGP occurs late in the evolution of the 
dense hadronic matter we must preserve the strangeness conservation 
criterion in
our fit. It will be of some considerable importance for this 
alternative study  to have a measure 
of the relative  abundance of kaons and hyperons. The issue is that all 
open-strangeness hadronic particles are now produced at the same
time in the evolution of the fireball and thus should definitively 
add up to give 
$\langle s-\bar s\rangle\simeq 0$. This is here a very strong constraint
as it in principle establishes a  consistency between different otherwise 
seemingly unrelated particle abundances. 
An experimental measure of the relative abundance strength is 
so far not given. However, we note that the NA49 spectra
\cite{NA49B} of kaons and hyperons have a slightly overlapping domain of 
$m_\bot$. We recall that the slopes are different, thus all we can do is to
try to combine the two  shapes, assuming continuity consistent with flow, 
and to estimate the relative normalization of both that would place
all experimental points on a common curve. We have carried out this procedure 
and obtained:
\begin{eqnarray}
\left.\frac{\Lambda}{K^0_s}\right|_{m_\bot}\simeq 6.2\pm1.5\,.
\end{eqnarray}
Note that there is a tacit presumption in our approach 
that a similar effective  $\Delta y$ interval
was used in both spectra.  We recall that this ratio
was $4.5\pm0.2$ in the S-W data \cite{WA85K}.
We will now fit altogether 9 data points: the 8 baryonic ratios and our
above  estimate.
We show the resulting fitted statistical parameters in the third line of
 table \ref{t1}, and note that this cold-QGP alternative has a 
very comparable statistical significance as the hot-QGP.
The  computed flow velocity at freeze-out is $v_f=0.51$c. 
This is just below the  relativistic sound velocity 
$v_s=1/\sqrt{3}=0.58$ which we have assumed in the calculation 
shown in Fig.\,\ref{gammTf}. In that figure, 
the cross to the left shows the 
result of the fit; allowing for potentially smaller expansion 
velocity and all the above discussed uncertainties in the 
computation, this result must also be seen as a very good agreement 
between the result of data fit and the theoretical calculation.
This also means that we cannot distinguish in the present data between 
early formation of strange antibaryons and an expansion model followed
by direct global hadronization.  However, we note that in table \ref{t2}
the resulting particle ratios shown in column four, under heading `cold'
are quite different from the other QGP fits and thus we should be able to 
distinguish the scenarios with better experimental data.

We will finally consider the possibility that 
the strange particle abundances could
be explained within a simple thermal model of confined, equilibrated 
particles, commonly called hadronic gas (HG) emanating at a late 
stage of fire evolution. We can now employ a minimal 
set of parameters  since assuming fully
equilibrated hadronic gas we have $\gamma_s=1,\,R_c^s=1$, and
we use strangeness conservation to evaluate the strangeness fugacity 
$\lambda_s$. We show the result of the fit in the last line 
of  table \ref{t1} and in particular we note:
\begin{eqnarray}
T_f&=&156\pm8.7\,\mbox{MeV},\rightarrow v_\bot\simeq 0.5\simeq v_s;\nonumber\\
\lambda_q&=&1.56\pm0.07, \ \rightarrow\  \lambda_s=1.14;\nonumber\\
\chi^2/9&=&0.84,\ \ \rightarrow \ \ \mbox{C.L.}>60\%\,.
\end{eqnarray}
We recall that the baryochemical potential is given in 
terms of $T$ and $\lambda_q$, $\mu_b=3T\ln \lambda_q$, and we 
find $\mu_b=205\pm10$ MeV in this hadronic gas condition. 
We note that while the quality of the fit has degraded, it 
still has considerable statistical significance.
The resulting particle ratios is  shown in last column in
table \ref{t2}. Looking at it, we see that it  is very hard 
with `naked' eye to give 
preference to any of the fits to the data, though this last one is
definitively less statistically significant than the other three.

\section*{\normalsize\bf Concluding Remarks} 

Clearly, the remarkable effect that we see in the data is that
within factor 1.4 all available Pb--Pb results can be explained in
terms of fully thermally and chemically equilibrated hadronic gas. 
If this is born out by more precise data, we will have a considerable
interpretation problem to solve: a back of envelope calculation 
shows that equilibration constants on HG are far too slow to
explain this effect, and thus either we err significantly in our
estimates about the evolution of HG or there is some 
chemical equilibration mechanism, such as QGP formation. If we 
apply QGP model to the data, we find a more statistically
significant fit, but cannot distinguish at present 
the type of  fireball evolution that is occurring. All this shows that 
we really need to have better data to make progress, and these 
should include not only ratios of strange particles but also
reflect on the global strangeness enhancement, 
and a measure of specific entropy per baryon which, as we did 
stress previously \cite{entropy} provides additional evidence for the
color bond breaking of hadrons as deconfinement sets in.

Our quite successful fits to the data  show
 that the freeze-out properties of strange
 baryons and antibaryons
point to a well localized, confined, thermally and 
nearly chemically 
equilibrated source, undergoing possibly a 
transverse expansion nearly with the velocity of sound
of relativistic matter. 
Some preference of the data points towards the QGP picture of 
the reaction, but assumption of an equilibrated HG also
explains the present day Pb--Pb data, and more and better data
is needed to resolve the issue. However, we have been able to
obtain here a rather specific information on the chemical 
conditions in the dense matter fireball formed in 
central Pb--Pb collisions; this information is 
of considerable relevance  for study of other phenomena.
We determined that quark fugacity is
$\lambda_q\simeq 1.53$ corresponding to 
$\lambda_b=\lambda_q^3\simeq 3.6$ (and thus
baryochemical potential $\mu_b=T\log \lambda_b=200$ MeV 
for `cold' QGP or HG freeze-out condition), and a strange
quark fugacity $\lambda_s\simeq 1.14$.

\subsection*{Acknowledgment}
 J.R. acknowledges partial support by  DOE, grant
               DE-FG03-95ER40937\,.

\end{document}